# Quasi-freestanding monolayer heterostructure of graphene and hexagonal boron nitride on Ir(111) with a chiral boundary

Mengxi Liu[1, †], Yuanchang Li[2, †], Pengcheng Chen[2], Jingyu Sun[1], Donglin Ma[1], Qiucheng Li[1], Teng Gao[1], Yabo Gao[1], Zhihai Cheng[2], Xiaohui Qiu[2], Ying Fang[2], Yanfeng Zhang[1,3,*], & Zhongfan Liu[1,*]

[1]Center for Nanochemistry (CNC), Beijing National Laboratory for Molecular Sciences, College of Chemistry and Molecular Engineering, Academy for Advanced Interdisciplinary Studies, Peking University, Beijing 100871, People's Republic of China

[2]National Center for Nanoscience and Technology, Chinese Academy of Sciences, Beijing 100190, People's Republic of China

[3]Department of Materials Science and Engineering, College of Engineering, Peking University, Beijing 100871, People's Republic of China

[†]These authors contributed equally to this work.

*Correspondence and requests for materials should be addressed to Y. F. Zhang & Z. F. Liu (E-mail: yanfengzhang@pku.edu.cn; zfliu@pku.edu.cn)

**Monolayer lateral heterostructure of graphene and hexagonal boron nitride (*h*-BNC) has attracted a growing attention mainly due to its tunable band-gap character and unique physical properties at interface. Hereby, we reported the first-time synthesis of a nearly freestanding *h*-BNC hybrid on a weakly coupled substrate of Ir (111), where graphene and *h*-BN possessing different surface heights and corrugations formed a perfect monolayer hybrid. With the aid of scanning tunneling microscopy/spectroscopy (STM/STS), we demonstrated that *h*-BN can patch alongside the boundary of pre-deposited graphene domains and vice versa to form a seamless monolayer hybrid, with the realization of predominant zigzag type chiral boundaries at the interface. Density-functional theory calculations and STM/STS measurements aided us to reveal that this interface between graphene and *h*-BN were atomically sharp in aspects of the chemical bonding as well as the local electronic property from both theoretical and experimental points of view.**

A long-existing challenge for graphene applications in electronics and photonics lies in the band-gap opening, which enables the modification of graphene into a semiconductor or even an insulator.[1-3] Latest theoretical calculations, pointing out that the lateral hybrid of graphene and hexagonal boron nitride (*h*-BNC) is semi-conductive with a band-gap of ~1eV, have inspired the exploration in the target synthesis of such material to probe its novel properties like magnetism.[4-10]

Experimentally, a pioneer work in this field has centered on the synthesis of randomly distributed graphene and *h*-BN domains on Cu foils, where a small band-gap opening (18 meV) was observed.[11] Several follow-up attempts on "patterned re-growth" of in-plane *h*-BNC hybrids have been made by virtue of patching graphene into pre-patterned monolayer *h*-BN (by means of photolithography), offering special insights into building-up atomically thin integrated circuitries and two-dimensional (2D) complex devices.[12,13] Very recently, the heteroepitaxial growth of *h*-BN from the fresh graphene domain edge leading to the formation of a continuous monolayer of *h*-BNC has also been well demonstrated.[14]

Although research has been flourished into this area, the reported protocols for the synthesis of in-plane *h*-BNC hybrid have always concerned with the usage of highly corrugated Cu foils as the growth substrate, which nevertheless prevented further chemical identification of the composite domains, the atomic lattice coherence at the linking edge, the edge type (armchair or zigzag), as well as the influence of *h*-BN to the electronic property of graphene at atomic scales in its as-grown state.[14-16] As a replacement, *in situ* growth of *h*-BNC on single crystalline substrates of Rh (111) and Ru (0001) were realized under ultra-high vacuum (UHV), which could benefit from the *in-situ* scanning tunneling microscopy/spectroscopy (STM/STS) study on the constitutional and morphological identifications and even the electronic property characterizations.[17-19] On Ru (0001), it was reported that *h*-BN and graphene evolved into *h*-BC$_2$N alloys at the interface regions due to the chemical reactions of ethylene and borazine precursor.[17,19] In contrast, on Rh (111), graphene and *h*-BN were suggested to patch seamlessly into an atomic layer, with a preferred zigzag edge (~77%) between the two analogues.[18] However, it is worth-noting that, either Rh (111) or Ru (0001) couples strongly with graphene by forming π-d orbital hybridization, leading to a downward shift (away from Fermi level) or even the breakage of the π bands of graphene.[20-23] Therefore, to better explore the growth features as well as examine the electronic characteristics of the *h*-BNC hybrid, a weakly coupled system would be highly demanded.

In this work, a sequential two-step growth method was designed to accomplish the synthesis of monolayer *h*-BNC hybrid on a weakly coupled substrate of Ir(111)[24] under UHV conditions. The linking edge of graphene and *h*-BN can be dictated to be mainly zigzag type (~95%), as characterized by high resolution (HR) STM. Intriguingly, the *h*-BNC hybrid on Ir (111) was of quasi-freestanding nature since *h*-BN and graphene presented its own intrinsic electronic feature within respective domain, whilst a sharp transition occurred at the interface of the two analogues with respect to the atomic arrangements as well as the electronic properties. It was further demonstrated herein that graphene and *h*-BN at different corrugations were able to form a perfect monolayer hybrid accompanied by evident strain relaxation routes, where the strain arising from the adlayer-substrate lattice mismatch can be released throughout wrinkle or crooked moiré formation.

**Graphene and *h*-BN on Ir(111).** The atomic-scale observation of the discrete features of graphene and *h*-BN on Ir (111) should be a prerequisite for a clear identification of the related hybrid material. Hereby, we managed to grow monolayer graphene and *h*-BN on Ir (111) and characterize their different morphologies. The STM images in Fig. 1a, b show that graphene on Ir (111) is featured by a periodic superstructure or graphene moiré pattern, arising from a coincidence lattice of 10×10 C-C/9×9 Ir (111) with a period of ~2.5 nm, as inferred from the atomically resolved STM image (Fig. 1c) and DFT calculation (Fig.1g), both of which display nearly the same moiré contrasts. An average distance of ~0.42 nm for graphene with regard to the Ir (111) plane, as well as a tiny layer corrugation of 0.02 nm (black curve in Fig. 1j), could be obtained by STM height profile analysis, which is in good agreement with the results gained from our theoretical calculations (black curve in Fig. 1i). It is worth-noting that the layer corrugation of graphene on Ir(111) is much smaller than that on Rh (111) (~0.16 nm)[18] or Ru (0001) (~0.15nm),[25] strongly indicative of a weak adlayer-substrate interaction between graphene and Ir (111).

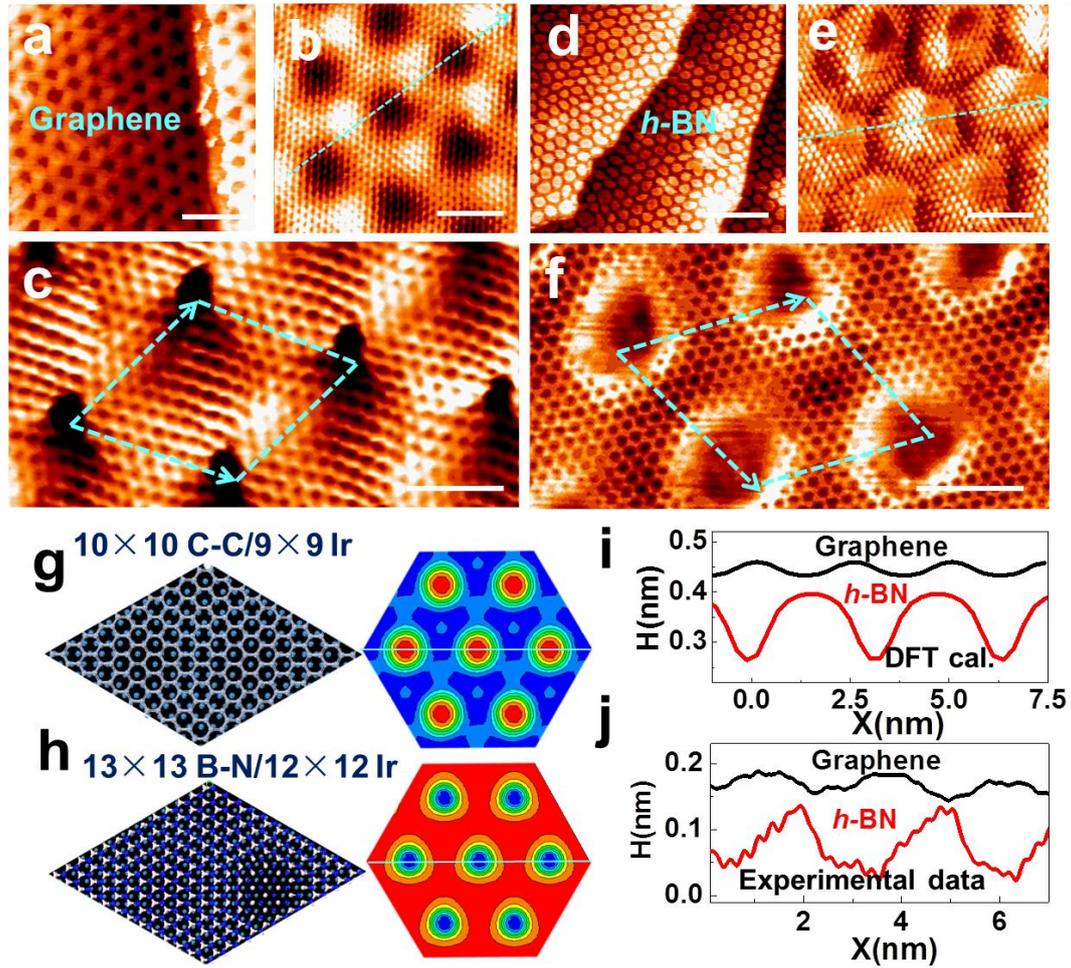

**Figure 1 UHV-STM observation and DFT simulation of pure graphene and *h*-BN monolayer on Ir (111).** (**a-c**) STM images of graphene at moiré and atomic scales. (a, b, c: $V_T$ = -0.138, -0.002, -0.002 V; $I_T$ = 5.361, 11.492, 4.874 nA, respectively) (**d-f**) STM images of *h*-BN at moiré and atomic scales. (d, e, f: $V_T$ = -0.037, -0.004, -0.002 V; $I_T$ = 1.284, 2.751, 52.267 nA, respectively) (**g, h**) DFT calculations of the surface corrugations and the adlayer-substrate distances of graphene and *h*-BN on Ir (111), respectively. (**i, j**) Comparison of the theoretical and experimental data of adlayer corrugations of graphene and *h*-BN on Ir (111), as derived from the cross-sectional views of the height mapping in (g, h) and the STM images in (b, e), respectively. Scale bars: (a) 8 nm; (b) 2.5 nm; (c) 1.5 nm; (d) 15 nm; (e) 3 nm; (f) 2 nm.

To the best of our knowledge, there has been no STM study so far with regard to *h*-BN/Ir (111) system, herein, our work reports the first-time STM characterization of *h*-BN grown on Ir (111), as depicted in Fig. 1d-f. *h*-BN on Ir (111) is featured by a periodic nanobubble structure at low magnifications (Fig. 1d, e) with a period of ~3.2 nm (larger than that of graphene on Ir (111) of ~2.5 nm). Zoom-in STM micrograph in Fig. 1f presents a coincidence lattice of 13×13 B-N/12×12 Ir(111), as displayed by the simulated graph in Fig. 1h. Likewise, DFT calculations indicate that *h*-BN is positioned ~0.38 nm away from the Ir(111) plane with a corrugation of ~0.14 nm (red curve in Fig. 1i), the value of which is much greater than that of graphene on Ir (111) (~0.02 nm). Although previous reports have suggested that the interaction between *h*-BN and Ir (111) was also quite weak,[26,27] it would still be interesting to see the coexistence of seamless patching of graphene and *h*-BN on Ir (111) plane, since we have just shown here that individual domains of graphene and *h*-BN possess distinct moiré roughness and adlayer-substrate distances on Ir (111).

**Quasi-freestanding *h*-BNC hybrid on Ir (111).** Surprisingly, the fabrication of h-BNC hybrids has been realized on Ir (111) by employing a specific two-step growth approach, which allowed direct control over the sequential patching of graphene and h-BN with distinct domain shapes (Fig. 2a and 2d). On one hand, as schematically illustrated in Fig. 2a, sub-monolayer graphene is deposited on Ir(111) using $C_2H_4$ as the carbon precursor, followed by patching *h*-BN into the remaining bare substrate by altering the precursor to boron nitride powder, giving rise to the formation of graphene embedded within *h*-BN structure (or G@*h*-BN hybrids). The surface feature of pre-formed graphene can be examined prior to the supply of *h*-BN precursor (Fig. 2b). It can be observed that hexagon-like graphene islands prevail after first step growth due to the intrinsic symmetry of graphene. In particular, the facts that (i) the edge direction of the graphene island stays in parallel with the direction of graphene moiré pattern (as shown in Fig. 2b), and (ii) the zigzag orientation of the atomic lattice is in line with the direction of moiré pattern (as indicated by the arrows in the atom-resolved STM image in Fig. 2b inset), substantiate that the near hexagonal graphene island is actually of zigzag termination. Notably, this relation is essential for labeling the edge type of as-synthesized *h*-BNC hybrids.

After the second step growth of *h*-BN, graphene and *h*-BN domains can be seamlessly patched together to exhibit "nanomesh" and "nanobubble" surface features with periods of ~2.5 nm and ~3.2 nm, respectively (Fig. 2c). Intriguingly, the pre-deposited graphene islands maintain the initial hexagonal shapes, the phenomenon of which is different from that occurred in the *h*-BNC growth on Ru (0001) by presenting transitional regions with a mixture of C, B, and N atoms,[17, 19] therefore suggesting that a sharp interface exists between graphene and *h*-BN on Ir (111).

On the other hand, *h*-BN embedded within graphene structure (or *h*-BN@G hybrids) can also be targetedly fabricated by sequentially growing *h*-BN and graphene on Ir (111), as illustrated in Fig. 2d. In this regard, the pre-grown *h*-BN island forms a triangular shape (Fig. 2e) with a preferred zigzag type edge (inset in Fig. 2e). It is also noticeable that the shape of the triangular *h*-BN island keeps almost unchanged after the second step growth of graphene (Fig. 2f), probably implying a clear interface between the two composite domains. X-ray photoelectron spectroscopy spectra of as-grown samples display B, C, and N 1s core level peaks, further verifying the formation of *h*-BNC hybrid structures in both scenarios (*Supplementary Information* Fig. S1).

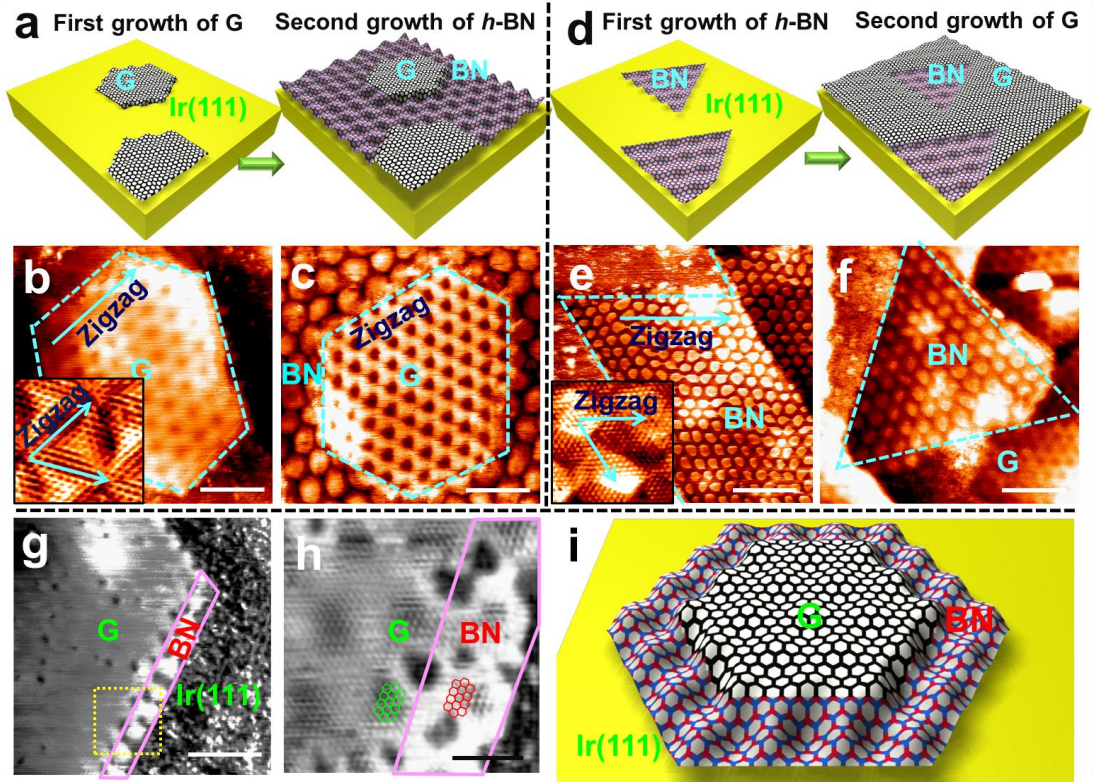

**Figure 2. Two-step patching growth of *h*-BNC hybrids.** (**a**) Schematic of the sequential growth steps for G@*h*-BN hybrid. (**b**) STM image of pre-deposited graphene islands on Ir (111) with the atom-resolved image (inset) showing the zigzag orientation of the domain. ($V_T$ = -0.194, $I_T$ = 2.501 nA; inset: $V_T$ = -0.002, $I_T$ = 40.953 nA) (**c**) STM image of G@*h*-BN hybrid. ($V_T$ = –0.002, $I_T$ = 1.879 nA) (**d**) Schematic of the sequential growth steps for *h*-BN@G hybrid. (**e**) STM image of the pre-deposited *h*-BN island on Ir (111) with the atom-resolved image (inset) showing the zigzag domain edge. ($V_T$ = -0.122, $I_T$ = 2.134 nA; inset: $V_T$ = -0.002, $I_T$ = 25.429 nA) (**f**) STM image of *h*-BN@G hybrid. ($V_T$ = -0.152, $I_T$ = 1.820 nA) (**g, h**) Sequential zoom-in image of *h*-BN moiré row growing along the edge of a graphene domain. (g, h: $V_T$ = -0.211, -0.021 V; $I_T$ = 1.820, 1.820 nA, respectively) (**i**) Schematic of the growth characteristics at the transition region. Scale bars: (b) 7 nm; (c) 8 nm; (e) 15 nm; (f) 10 nm; (g)10 nm; (h) 3nm.

**Zigzag interface boundaries of the *h*-BNC hybrid.** To gain further understanding on the formation process of *h*-BNC hybrid, the initial patching behavior of *h*-BN onto a pre-deposited hexagonal graphene domain was probed (Fig. 2g and h), where a

*h*-BN moiré row grows preferentially along the graphene edge. This perferred formation of hybrid structures over individually separated domains could be attributed to the possible existence of dangling bond at fresh graphene edges, which could facilitate the sequential growth of *h*-BN. Moreover, from the energetic point of view, the average binding energy of the *h*-BNC hybrid is proposed to be much higher than that of pure graphene or *h*-BN, as previously reported for *h*-BNC growth on Rh(111).[18] Notably, even at sub-monolayer coverage, a sharp interface between graphene and *h*-BN can still be clearly identified with the aid of their different moiré contrasts, where the scheme in Fig. 2i presents a vivid visualization for the unique feature at the graphene-BN interface region.

To further probe this sharp interface transition and the interface edge type, atom-resolved STM images have been captured at the regions of interest (Fig. 3a-d), showing that *h*-BN grows strictly along the fresh edges of pre-existing graphene with atomic lattice coherence to form a continuous monolayer. However, the two analogues can be clearly distinguished from each other in STM views due to their different moiré contrasts. More importantly, a zigzag-type boundary linking graphene and *h*-BN can be clearly observed from the atom-resolved images, as evidently marked by the fitted net models ( more STM images of zigzag linking type seeing *Supplementary Information* Fig. S2-S8). It is noted that this zigzag-type chiral boundary is universal in the present system (> 95%), which can also be identified from moiré-scale STM images when graphene and *h*-BN are patching together to form a line shape interface, as illustrated in Fig, 2 as well as in the previous work of *h*-BNC on Rh(111).[18] Therefore, our work herein highlights a simple two-step growth method to create *h*-BNC hybrid with spontaneous formation of zigzag boundaries. This edge termination has been suggested to present intriguing properties such as half metallicity and spin polarizations from both experimental andtheoretical points of view.[7, 9, 10]

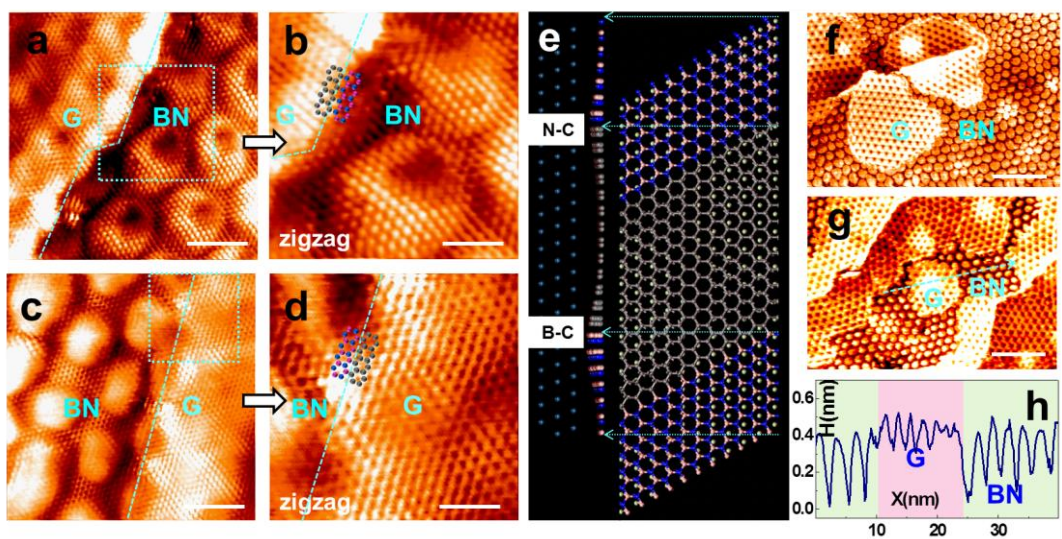

**Figure 3. The edge type identification by atom-resolved STM investigations.** (a-d) Atom-resolved STM images at the interface boundaries showing predominant zigzag edges. (a, b, c, d: $V_T$ = -0.004, -0.004, -0.002, -0.002 V; $I_T$ = 2.750, 2.750, 1.655, 1.504 nA, respectively) (**e**) DFT calculation of the corrugation at the interface regions of B-C and N-C linkage. (**f, g**) STM images of monolayer hybrid showing clear contrast differences in the regions of graphene, *h*-BN and their interface. (f, g: $V_T$ = -0.002, -0.002V; $I_T$ = 2.274, 3.870 nA, respectively). (**h**) STM height profile recorded along the arrow in (g). Scale bars: (a) 3 nm; (b) 1.2 nm; (c) 2.5 nm; (d) 1.2 nm; (f) 15 nm; (g) 20 nm.

DFT calculations of the corrugation at the interface regions also verify that a seamless zigzag edge with either B-C or N-C linkage could form on Ir (111), with the B-C edge pulling down more seriously than that of the N-C edge (Fig. 3e). The B-C linkage is therefore considered to be more stable than that of the N-C linkage on Ir (111), consistent with the observation that the B-C edge possessing a larger binding energy than that of the N-C edge within the *h*-BNC hybrid on Rh(111).[18] It is noted that this interface height undulation is not revealed because of the resolution limit of STM, detailed atomic force microscopy study aiming to provide a clearer identification of the interface lattice arrangement as well as the height undulation is still ongoing. However, from the large-scale STM image with both analogues situating on the same terrace of Ir (111), one can clearly identify the contrast differences between graphene, *h*-BN, and their interface (Fig. 3f and g), where graphene adopts a brighter contrast (~0.1 nm higherthan that of *h*-BN) with smaller surface undulation (~0.1nm for graphene *versus* ~0.4 nm for *h*-BN), consistent with the results obtained from the DFT calculations.

The above results indicate that graphene and *h*-BN can overcome the differences in height and moiré period on Ir (111) to evolve a seamless monolayer hybrid with a predominant zigzag linking edge. In this regard, we can infer that the patching window of graphene and *h*-BN can be rather wide on Ir (111), which differs dramatically from those in previously reported *h*-BNC system on Ru (0001) and Rh (111).[17-19]

**Electronic properties at the interface of *h*-BNC.** Along with the information regarding the sharp transition of chemical bonding at the patching interface, it would be meaningful to know whether a sharp electronic state transition could also occur at the interface, as well as whether the patching of *h*-BN could affect the electronic property of graphene (Fig. 4a). To answer these questions, typical dI/dV curves of monolayer graphene and *h*-BN on Ir (111) were captured, respectively (Fig. 4b). It is clear to see that the Dirac point of graphene/Ir (111) locates at 0 eV, indicating almost no doping effect of Ir(111) on the as-grown graphene, reconfirming a weak interaction between graphene and Ir (111) substrate. Likewise, a nominal band-gap of ~5 eV is visible on as-grown monolayer *h*-BN/Ir (111), the value of which is quite close to that of bulk *h*-BN (5.2-5.7 eV), once again indicative of weak adlayer-substrate interaction.

Sequential line scan (as indicated in Fig. 4a) dI/dV data across a zigzag patching edge from graphene to *h*-BN regions were obtained,where each spectrum labled with a letter (Fig. 4c) corresponds to an ascertained position in Fig. 4a. It can be observed that, on pure graphene region (~20 nm away from the edge), the spectrum appears to be a V-type shape with the dirac point locating at ~0 eV (spectrum A in Fig. 4c), which is typicl for freesanding graphene; on pure *h*-BN region (~20 nm away from the edge), the dI/dV signal presents a flat background (differential conductance close to zero) from -0.8 eV to 0.8 eV (spectrum F in Fig. 4c), suggesting the insulating nature of *h*-BN when synthesized on a weakly coupled susbtrate of Ir (111).

Intriguingly, when the scan approaches the linking edge (from graphene side), a prominent broadening of the V type dI/dV signals occurs (spectrum B in Fig. 4c), with the indication that the differential conductance is completely suppressed at the interface regions(Fig. S9). This shrinkage of the graphene conductance could be attributed to either the edge effect of graphene island or the large band-gap value of *h*-BN as a potential barrier of the electron from graphene.[28] Although further study from both experimental and theoritical aspects would be helpful to reveal more detailed reasons, it can be concluded from this investigation that, a sharp electronic state transition indeed occurs at the patching interface of graphene and *h*-BN.

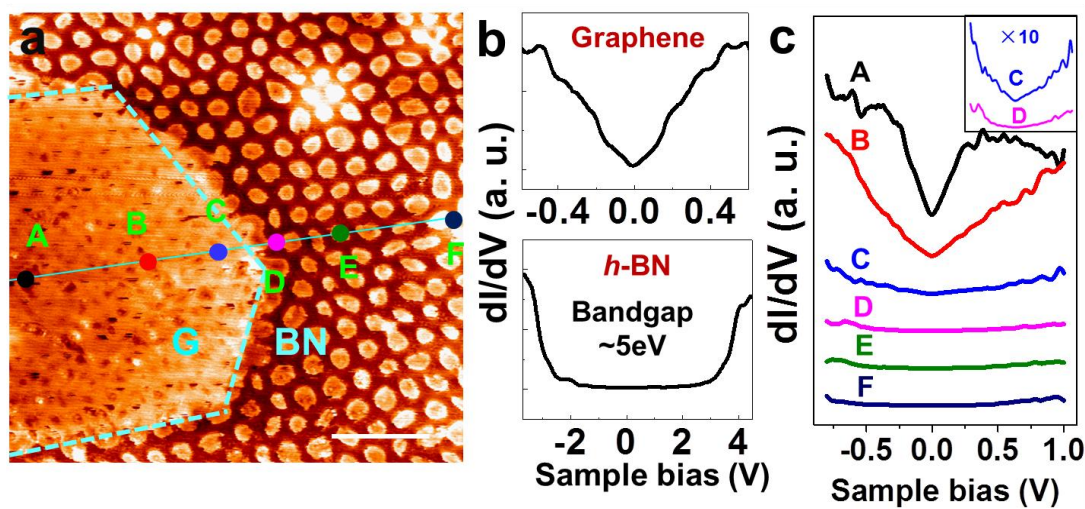

**Figure 4. Sharp transition of the electronic state at the interface of graphene and *h*-BN.** (a) STM image of G@*h*-BN hybrid. ($V_T = -0.183$, $I_T = 2.134$ nA). Scale bar: 10 nm. (**b**) Typical dI/dV curve of monolayer graphene and *h*-BN on Ir (111), respectively. (**c**) Single point dI/dV curves measured along the points marked in (a).

**Strain releasing of *h*-BNC hybrid on Ir(111).** In the present study, the observation that graphene and *h*-BN can be patched seamlessly to produce a monolayer in-plane hybrid lead us to probe the strain effect, where the strain is released in the form of defect formation. It is found that, at the graphene areas within the hybrid, graphene wrinkles (or nanobubbles) can be observed occasionally on the step edges of Ir (111) (Fig. 5a-c) with an apparent height of ~1.3 nm, which is much lower than that of the wrinkles on CVD graphene (~3-10 nm).[15,29] On the top of bumped wrinkle (Fig. 5c), a honeycomb graphene lattice can be resolved clearly, verifying the formation of graphene wrinkles. The corrugated graphene wrinkles could be effective in tuning the electronic states of graphene *via* the creation of a pseudo-magnetic field.[30] Another route for strain relaxation is proposed to be throughout the irregular shaped or crooked superstructure formation at the interface (as schematically shown in Fig. 5d), which, however, exerts no influence in the seamlessness of the hybrid, where graphene and *h*-BN still maintain continuous at the atomic scale at their patching interface with a series of short zigzag type edges (Fig. 5e, f, and large-scale STM image in Fig. S10).

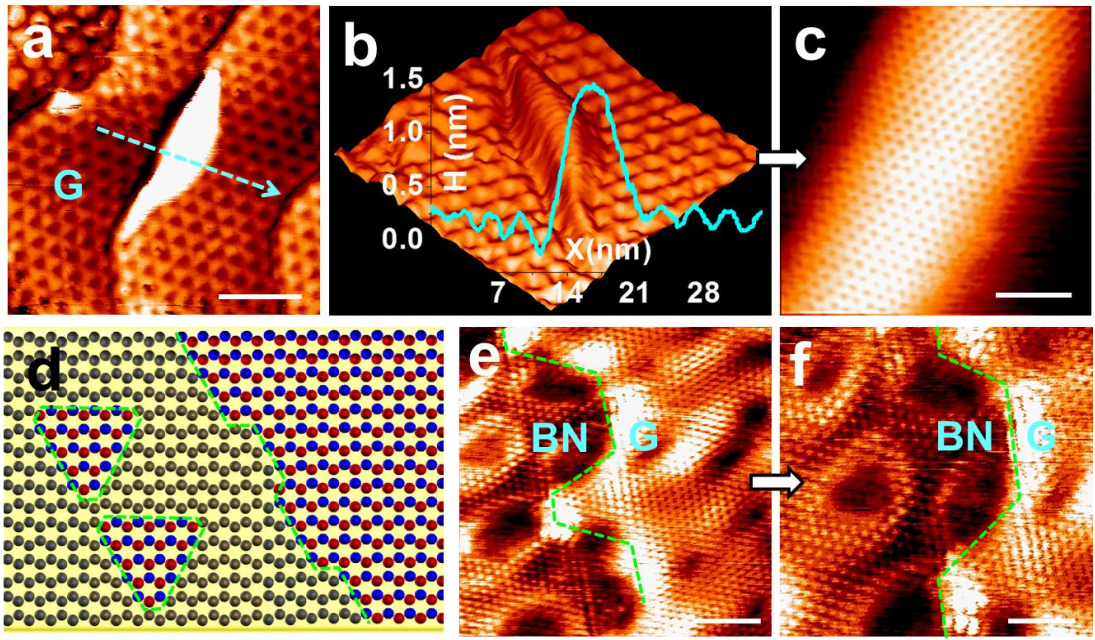

**Figure 5. Pathways for strain releasing.** (**a-c**) Large-scale and atom-resolved STM images of wrinkle and crooked moiré formation. (a, c: $V_T$ = -0.016, -0.016 V; $I_T$ = 2.860, 2.860 nA, respectively) (**d**) Schematic of irregular shape moiré formation on the patching interface linking graphene and *h*-BN. (**e, f**) Atomic-scale STM images on the disordered linking edges. (e, f: $V_T$ = -0.007, -0.007 V; $I_T$ = 1.655, 1.655 nA, respectively) Scale bars: (a) 10 nm; (c) 1 nm; (e) 2 nm; (f) 1.2 nm.

In conclusion, we have accomplished the first-time synthesis of a nearly freestanding monolayer *h*-BNC hybrid on a weakly coupled substrate of Ir (111), regardless of the disparate moiré period, moiré contrast, adlayer-substrate distance, and layer corrugation of graphene and *h*-BN on this substrate. The interface between graphene and *h*-BN has shown to be atomically sharp in the aspect of the chemical bonding as well as the local electronic property. More importantly, a seamless interface of graphene and *h*-BN with predominant zigzag type chiral boundaries can be realized *via* a two-step patching growth method, as convinced by our atom-resolved STM examinations of the growth process as well as the produced hybrid. Our study is of fundamental importance for understanding the controlled growth under UHV and the interface morphology, chemistry, and electronics at the atomic level of the *h*-BNC monolayer in-plane hybrids, offering special insights into the construction of different types of monolayer hybrids with intriguing properties.

**Methods**

**Preparation of *h*-BNC hybrids.** The Ir (111) single crystal (MaTeck GmbH, 99.99 % purity) was cleaned by repeated cycles of $Ar^+$ sputtering at room temperature [P(Ar) = $6 \times 10^{-6}$ mbar, 1.2 keV], pre-heating to ~1000 K, and annealing in an oxygen atmosphere [P($O_2$) = $1 \times 10^{-7}$ mbar] at ~900 K. Graphene and *h*-BN were fabricated by the thermal decomposition of ethylene ($C_2H_4$) gas and borane amines ($BH_3NH_3$) at 800-900 K, respectively.

**STM/STS Characterizations.** Omicron VT-STM and LT-STM/STS systems were used for the fabrication and characterization of samples with a base pressure better than $10^{-10}$ mbar. All of the STM data were captured under a constant current mode with the sample kept at room temperature. The local differential conductance (dI/dV) spectra were measured at 77 K by recording the output of a lock-in system with the manually-disabled feedback loop. A modulation signal of 5 mV, 952 Hz was selected under a tunneling condition of 1 V, 20 pA.

**DFT Calculations.** We performed the density-functional theory (DFT) calculations using the Vienna *ab-initio* simulation package (VASP)[31] with the Perdew-Burke-Ernzerhof[32] exchange-correlation. The projector-augmented wave[33] potential with a cutoff energy of 400 eV was used to describe the electron-ion interaction. In order to consider the interaction between graphene/*h*-BN and Ir accurately, van der Waals (vdW) interaction was included through the DFT-D2 method.[34] The parameters for Ir were used as that of Au, namely, 40.62 J nm$^6$/mol for the dispersion coefficient ($C_6$), and 1.772 Å for the vdW radius ($R_0$).[35] Only Γ point of the Brillouin zone was used to sample due to the large supercell. All the geometries were optimized without any symmetry constraint until the residual force on each atom was less than 0.06 eV/Å.

**References**

1. Novoselov, K. *et al.* Two-dimensional gas of massless Dirac fermions in graphene. *Nature* **438,** 197-200 (2005).

2. Geim, A. K.; Novoselov, K. S. The rise of graphene. *Nat. Mater.* **6,** 183-191 (2007).

3. Meyer, J. C.; Geim, A.; Katsnelson, M.; Novoselov, K.; Booth, T.; Roth, S. The structure of suspended graphene sheets. *Nature* **446,** 60-63 (2007).

4. da Rocha Martins, J.; Chacham, H. Disorder and Segregation in B− C− N Graphene-Type Layers and Nanotubes: Tuning the Band Gap. *ACS Nano* **5,** 385-393 (2010).

5. Shinde, P. P.; Kumar, V. Direct band gap opening in graphene by BN doping: Ab initio calculations. *Phys. Rev. B* **84,** 125401 (2011).

6. Zhao, R.; Wang, J.; Yang, M.; Liu, Z.; Liu, Z. BN-Embedded Graphene with a Ubiquitous Gap Opening. *J. Phys. Chem. C* **116,** 21098-21103 (2012).

7. Ramasubramaniam, A.; Naveh, D. Carrier-induced antiferromagnet of graphene islands embedded in hexagonal boron nitride. *Phys.Rev. B* **84,** 075405 (2011).

8. Bhowmick, S.; Singh, A. K.; Yakobson, B. I. Quantum dots and nanoroads of graphene embedded in hexagonal boron nitride. *J. Phys. Chem. C* **115,** 9889-9893 (2011).

9. Jiang, J.-W.; Wang, J.-S.; Wang, B.-S. Minimum thermal conductance in

graphene and boron nitride superlattice. *Appl. Phys. Lett.* **99,** 043103-043109 (2011).

10. Pruneda, J. Origin of half-semimetallicity induced at interfaces of C-BN heterostructures. *Phys. Rev. B* **81,** 161409 (2010).

11. Ci, L.*et al.* Atomic layers of hybridized boron nitride and graphene domains. *Nat.Mater.* **9,** 430-435 (2010).

12. Liu, Z.*et al.*In-plane heterostructures of graphene and hexagonal boron nitride with controlled domain sizes. *Nat.Nanotechnol.* **8,** 119-124 (2013).

13. Levendorf, M. P.*et al.*Graphene and boron nitride lateral heterostructures for atomically thin circuitry. *Nature* **488,** 627-632 (2012).

14. Liu, L.*et al.*Heteroepitaxial Growth of Two-Dimensional Hexagonal Boron Nitride Templated by Graphene Edges. *Science* **343,** 163-167 (2014).

15. Zhang, Y.*et al.* Defect-like structures of graphene on copper foils for strain relief investigated by high-resolution scanning tunneling microscopy. *ACS Nano* **5,** 4014-4022 (2011).

16. Cho, J.*et al*. Atomic-scale investigation of graphene grown on Cu foil and the effects of thermal annealing. *ACS Nano* **5,** 3607-3613 (2011).

17. Sutter, P.; Cortes, R.; Lahiri, J.; Sutter, E. Interface Formation in Monolayer Graphene-Boron Nitride Heterostructures. *Nano Lett.* **12,** 4869-4874 (2012).

18. Gao, Y.*et al.*Towards Single-layer Uniform Hexagonal Boron Nitride–Graphene Patchworks with Zigzag Linking Edges. *Nano Lett.* **13,** 3439-3443 (2013).

19. Lu, J.*et al*. Order–disorder transition in a two-dimensional boron–carbon–nitride alloy. *Nat.Commun.* **4,** 2681-2687 (2013).

20. Sutter, P.; Sadowski, J. T.; Sutter, E. A. Chemistry under Cover: Tuning Metal−Graphene Interaction by Reactive Intercalation. *J. Am. Chem. Soc.* **132,** 8175-8179 (2010).

21. de Parga, A. L. V. *et al*. Periodically rippled graphene: growth and spatially resolved electronic structure. *Phys. Rev. Lett.*, **100**, 056807 (2008).

22. Garnica M. *et al*. Long-range magnetic order in a purely organic 2D layer adsorbed on epitaxial graphene. *Nat. Phys.*, **9**, 368-374 (2013).

23. Wang, B.; Caffio, M.; Bromley, C.; Früchtl, H.; Schaub, R. Coupling epitaxy,

chemical bonding, and work function at the local scale in transition metal-supported graphene. *ACS Nano* **4,** 5773-5782 (2010).

24. Usachov, D. *et al.* Experimental and computational insight into the properties of the lattice-mismatched structures: Monolayers of *h*-BN and graphene on Ir (111). *Phys. Rev. B* **86,** 155151 (2012).

25. Martoccia, D. *et al.* Graphene on Ru (0001): a 25 × 25 supercell. *Phys.Rev.Lett.* **101,** 126102 (2008).

26. Laskowski, R.; Blaha, P. Ab initio study of *h*-BN nanomeshes on Ru (001), Rh (111), and Pt (111). *Phys. Rev. B* **81,** 075418 (2010).

27. Laskowski, R.; Blaha, P.; Schwarz, K. Bonding of hexagonal BN to transition metal surfaces: An ab initio density-functional theory study. *Phys. Rev. B* **78,** 045409 (2008).

28. Altenburg, S. J., Kröger, J., Wehling, T. O., Sache, B., Lichtenstein, A. I., and Berndt, R. Local gating of an Ir (111) surface resonance by graphene islands[J]. *Phys. Rev. Lett.* **108**, 206805 (2012).

29 Liu, M. *et al.* Thinning Segregated Graphene Layers on High Carbon Solubility Substrates of Rhodium Foils by Tuning the Quenching Process. *ACS Nano* **6,** 10581-10589 (2012).

30. Levy, N. *et al.* Strain-induced pseudo–magnetic fields greater than 300 tesla in graphene nanobubbles. *Science* **329,** 544-547 (2010).

31. Kresse, G.; Hafner, J. Ab initio molecular dynamics for liquid metals. *Phys. Rev. B* **47,** 558 (1993).

32. Perdew, J. P.; Burke, K.; Ernzerhof, M. Generalized gradient approximation made simple. *Phys.Rev.Lett.* **77,** 3865 (1996).

33. Blöchl, P. E. Projector augmented-wave method. *Phys. Rev. B* **50,** 17953 (1994).

34. Grimme, S. Semiempirical GGA‐type density functional constructed with a long‐range dispersion correction. *J.Comput.Chem.* **27,** 1787-1799 (2006).

35. Medeiros, P. V.; Gueorguiev, G. K.; Stafström, S. Benzene, coronene, and circumcoronene adsorbed on gold, and a gold cluster adsorbed on graphene: Structural and electronic properties. *Phys. Rev. B* **85,** 205423 (2012).

## Acknowledgements

This work was financially supported by the National Natural Science Foundation of China (Grants 51222201, 51290272, 11304053), the Ministry of Science and Technology of China (Grants 2011CB921903, 2012CB921404, 2012CB933404, 2013CB932603), and the Foundation for Innovative Research Groups of the National Natural Science Foundation of China (Grant 51121091).

## Author contributions

M.X.L., Y.F.Z. and Z.F.L. conceived the original idea. M.X.L. carried out the synthesis and STM study of *h*-BNC heterostructure. Y.C.L. provided the DFT calculations. M.X.L. and P.C.C. performed the STS measurements. T.G. and D.L.M. designed and drew the sketch images. M.X.L., J.Y.S., Y.F.Z. and Z.F.L. wrote the manuscript. All the authors discussed and commented on the manuscript.

## Competing financial interests

The authors declare no competing financial interests.